\documentclass[Physsubmission, Phys]{SciPost}

\pdfoutput=1

\hypersetup{
    colorlinks,
    linkcolor={red!50!black},
    citecolor={blue!50!black},
    urlcolor={blue!80!black}
}

\usepackage[bitstream-charter]{mathdesign}
\DeclareSymbolFont{usualmathcal}{OMS}{cmsy}{m}{n}
\DeclareSymbolFontAlphabet{\mathcal}{usualmathcal}

\usepackage{verbatim}

\begin{document}

\begin{center}{\Large \textbf{\boldmath 
  $\mathcal{O}(\alpha_s^3)$ corrections to semileptonic $b \to c$ decays
  in the heavy daughter approximation
}}\end{center}

\begin{center}
Kay Sch{\"o}nwald\textsuperscript{1$\star$}
\end{center}

\begin{center}
{\bf 1} Institut f{\"u}r Theoretische Teilchenphysik, Karlsruhe Institute of Technology (KIT), \\ 76128 Karlsruhe, Germany
\\
* kay.schoenwald@kit.edu
\end{center}

\begin{center}
\today
\end{center}

\definecolor{palegray}{gray}{0.95}
\begin{center}
\colorbox{palegray}{
  \begin{tabular}{rr}
  \begin{minipage}{0.1\textwidth}
    \includegraphics[width=35mm]{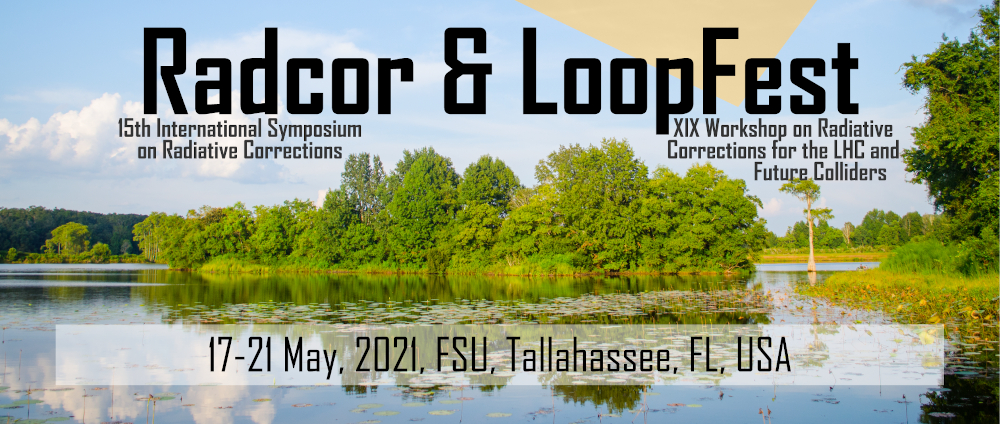}
  \end{minipage}
  &
  \begin{minipage}{0.85\textwidth}
    \begin{center}
    {\it 15th International Symposium on Radiative Corrections: \\Applications of Quantum Field Theory to Phenomenology,}\\
    {\it FSU, Tallahasse, FL, USA, 17-21 May 2021} \\
    \doi{10.21468/SciPostPhysProc}\\
    \end{center}
  \end{minipage}
\end{tabular}
}
\end{center}

\section*{Abstract}
{\bf \boldmath
We present our recent calculation of the third order
corrections to the semileptonic $b \to c$ and the muon decays.
The calculation has been performed in an expansion around the
limit $m_c \sim m_b$, but shows decent convergence even for $m_c=0$ from
which the contribution to the muon decay can be extracted.
For the semileptonic $b \to c$ decay we find large perturbative corrections
in the on-shell scheme which can be significantly reduced by changing
to the kinetic scheme for the bottom quark mass.
These results are important input for the inclusive determination
of $|V_{cb}|$ and the Fermi coupling constant $G_F$.
}

\section{Introduction}
\label{sec:1}
The Cabbibo-Kobayashi-Maskawa (CKM) matrix elements are fundamental constants 
in the Standard Model (SM) which describe the flavor mixing
in the quark sector and provide the only source of charge-parity (CP)
violation.
It is therefore important to determine these constants precisely.
One way to determine the CKM matrix elements $|V_{ub}|$ and $|V_{cb}|$ 
are inclusive semileptonic $B$ meson decays 
$B \to X_{c(u)} \ell \overline{\nu}$ using 
global fits to the experimental values of the semileptonic decay 
widths and moments of kinematical distributions
\cite{Bauer:2004ve,Gambino:2013rza,Alberti:2014yda,Gambino:2016jkc,Bordone:2021oof}.
Here, the presence of the heavy bottom quark allows to describe the 
decay in the heavy quark effective theory (HQET), where the decay 
rate can be given in an expansion in the strong coupling constant
$\alpha_s$ and in inverse powers of the heavy quark mass $1/m_b$.
The leading order in $1/m_b$ is given by the free quark decay 
$b \to c \ell \overline{\nu}$ which had been known up to 
$\mathcal{O}(\alpha_s^2)$ 
\cite{Luke:1994yc,Melnikov:2008qs,Pak:2008qt}
together with leading terms in the 
large $\beta_0$ approximation to higher orders
\cite{Ball:1995wa}.
Higher terms in the $1/m_b$ expansion are obtained from higher-dimensional
operators in the HQE.
In these proceedings we review the calculation of
the semileptonic decay rate at leading order in $1/m_b$ to order 
$\alpha_s^3$ obtained in Ref.~\cite{Fael:2020tow} and report on recent progress on 
the extension of the calculation to inclusive moments.

\newpage

\section{Calculation}
\label{sec:2}

\begin{figure}[t]
  \centering
  \begin{tabular}{ccc}
  \begin{tabular}{ccc}
    \raisebox{2.2em}{\includegraphics[width=0.25\textwidth]{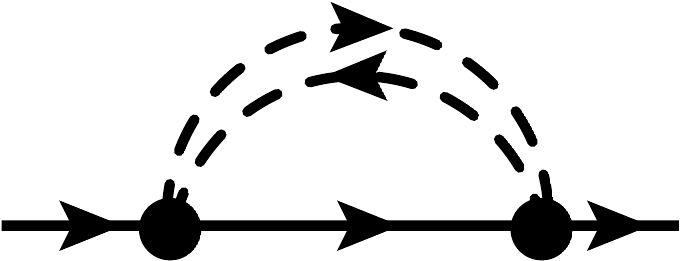}} &
    \includegraphics[width=0.25\textwidth]{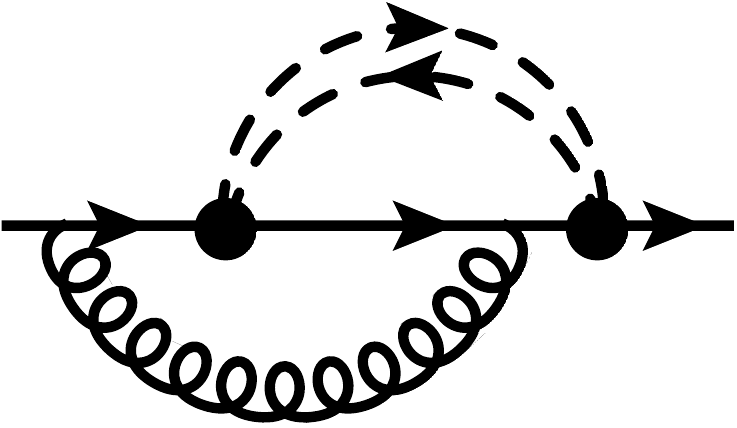} &
    \includegraphics[width=0.25\textwidth]{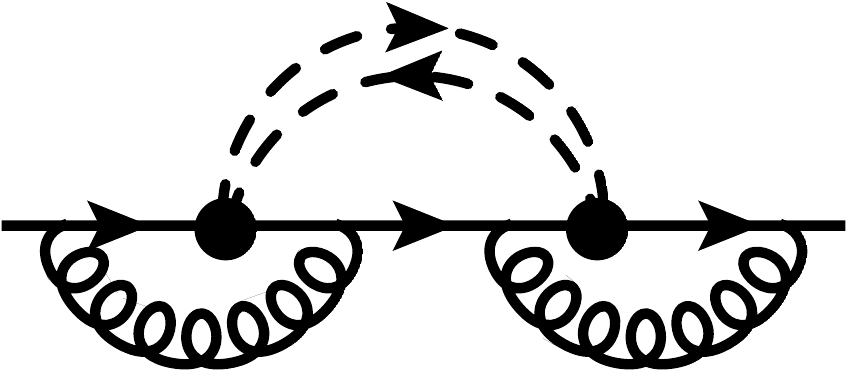} \\
    (a) & (b) & (c) \\
    \includegraphics[width=0.25\textwidth]{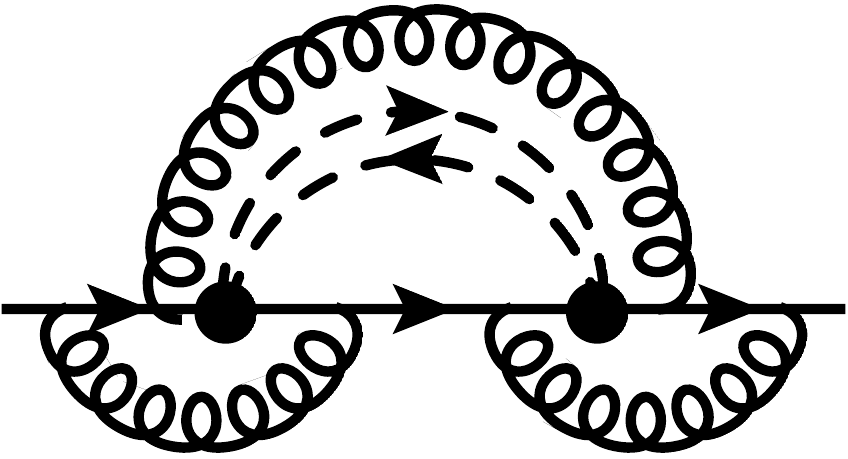} &
    \raisebox{0.3em}{\includegraphics[width=0.25\textwidth]{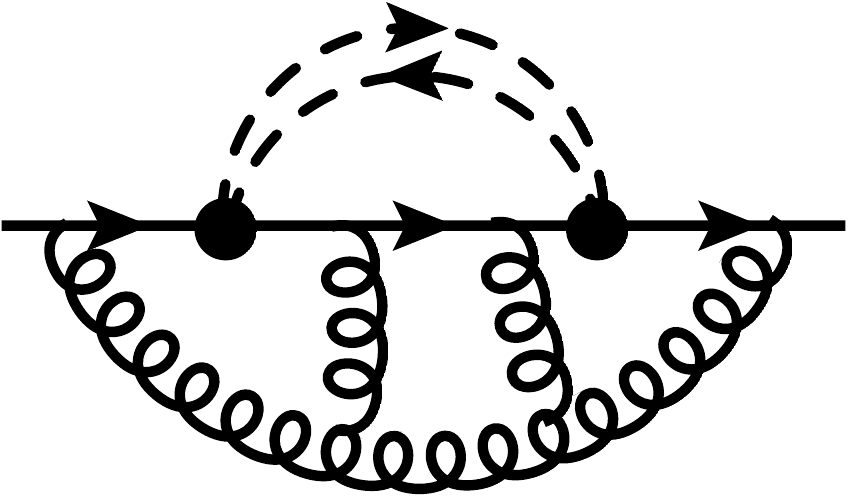}} &
    \includegraphics[width=0.25\textwidth]{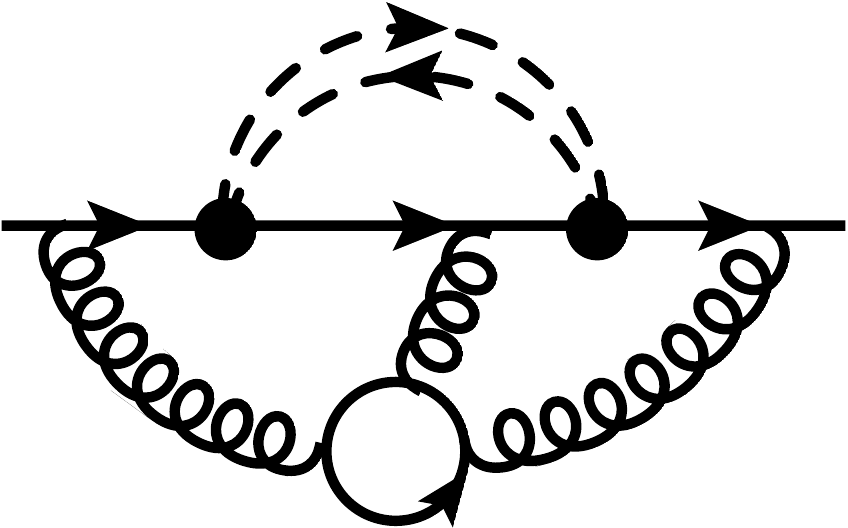} \\
    (d) & (e) & (f) \\
  \end{tabular}
  \end{tabular}
  \caption{Sample Feynman diagrams which contribute to the
    forward scattering amplitude of a bottom quark at LO (a), NLO (b), NNLO
    (c) and N$^3$LO (d-f). Straight, curly and dashed lines represent quarks,
    gluons and leptons, respectively. The weak interaction mediated by the $W$
    boson is shown as a blob.}
    \label{fig::diag}
\end{figure}

We compute the process
\begin{eqnarray}
  b(p) \to X_c(p_x)l(p_l) \overline{\nu}(p_\nu) ,
\end{eqnarray}
where $X_c$ is a state containing charm, light quarks and gluons.
The calculation is based on the optical theorem, this means 
we have to compute the imaginary part of 5-loop forward 
scattering diagrams.
Some example diagrams are given in Fig.~\ref{fig::diag}.
The total leptonic momentum is given by $q=p_\ell + p_\nu$.
For the global fits not only the total decay width but also 
moments of kinematic distributions are used.
In the following we will focus on the recently proposed 
$q^2$-moments \cite{Fael:2018vsp} which are defined by 
\begin{eqnarray}
  Q_i &=& \int {\rm d}q^2 (q^2)^i \frac{{\rm d} \Gamma}{{\rm d} q^2}
  ~.
\end{eqnarray}
Note that $Q_0$ corresponds to the total decay rate $\Gamma$. 
Inclusively these moments, including the total rate $\Gamma$, can be
computed from the imaginary part of the forward scattering diagrams 
by multiplying the integrands of the individual Feynman diagrams with 
the appropriate power of $q^2$ and then integrate.
For global fits, moments of the total hadronic invariant mass
and the charged lepton energy have been used.

Since an analytic calculation retaining the full dependence 
on the charm and bottom mass seems out of reach, we compute 
the diagrams in an asymptotic expansion around 
\begin{eqnarray}
  \delta &=& 1 - \frac{m_c}{m_b} \approx 0.7
  ~.
\end{eqnarray}
Although this limit seems unnatural for the physical values 
of the charm and bottom quark masses it has been shown 
at $\mathcal{O}(\alpha_s^2)$ in 
Ref.~\cite{Dowling:2008mc}
that this expansion converges quite 
fast at the physical point and can even be extended to $\delta \to 1$.
Furthermore, this limit has a couple of technical advantages:
\begin{itemize}
  \item To calculate the asymptotic expansion we 
  use the method of regions \cite{Beneke:1997zp}.
  In the limit $\delta \to 0$ the leptonic momentum has to 
  be ultrasoft $q \sim \delta \cdot m_b$. The number of regions 
  to be considered is therefore reduced. 
  \item When performing the $\delta$-expansion the leptonic 
  system completely factorizes and can be integrated out without 
  any IBP reduction. In the end, we are therefore 
  left with 3-loop integrals, although starting from 5 loops.
\end{itemize}
The asymptotic expansion has been implemented in dedicated 
{\tt FORM} \cite{Ruijl:2017dtg}
routines and we made use of the program {\tt LIMIT}
\cite{Herren:2020ccq}
to implement partial fractioning and the minimization of
topologies.
In the end we are left with integral families where either 
all loop momenta are hard or ultrasoft.
The first class of integrals corresponds to on-shell propagator
integrals which are well studied in the literature 
\cite{Laporta:1996mq,Melnikov:2000zc,Lee:2010ik}.
The second class has recently be considered for the relation 
between the kinetic and on-shell mass up to $\mathcal{O}(\alpha_s^3)$
\cite{Fael:2020iea,Fael:2020njb},
Here, the relevant master integrals are given 
to the necessary order in $\epsilon$ needed for the present calculation. 
Due to the large expansion depth of our calculation, we aim at 8 terms of 
the $\delta$-expansion, 
huge intermediate expressions of $\mathcal{O}(100\text{GB})$ had to be 
handled and $\mathcal{O}(10^{7})$
scalar integrals with positive or negative indices up to $12$
had to be reduced.
For this task we used {\tt FIRE} \cite{Smirnov:2019qkx} in combination with 
{\tt LiteRed} \cite{Lee:2012cn,Lee:2013mka}.\footnote{We thank A. Smirnov 
for providing a private version of {\tt Fire} which was essential for the 
reduction.} 

\section{Results}
\label{sec:3}
We parametrize the total decay rate as 
\begin{eqnarray}
  \Gamma &=& \Gamma_0 \left( X_0 
  + C_F \sum\limits_{i=1}^\infty \left( \frac{\alpha_s}{\pi} \right)^i X_i
  \right)~,  
\end{eqnarray}
with $\Gamma_0=A_{\rm ew} G_F^2 |V_{cb}|^2 m_b^5/(192 \pi^3)$,
$\alpha_s \equiv \alpha_s^{(5)}(\mu_s)$, 
$X_0=1-8\rho^2-12\rho^4\ln(\rho^2)+8\rho^6-\rho^8$,
$\rho = m_c/m_b$ and 
$A_{\rm ew}=1.014$ is the leading electroweak correction \cite{Sirlin:1981ie}.
Our result for the total decay rate at $\mathcal{O}(\alpha_s^3)$ reads
\begin{eqnarray}
  X_3 &=& 
  \delta^5 
  \biggl(
     \frac{266929}{810}
    -\frac{5248 a_4}{27}
    +\frac{2186 \pi ^2 \zeta _3}{45}
    -\frac{4094 \zeta_3}{45}
    -\frac{1544 \zeta _5}{9}
    -\frac{656 l_2^4}{81}
    +\frac{1336}{405} \pi ^2 l_2^2
    \nonumber \\ &&
    +\frac{44888\pi ^2 l_2}{135}
    -\frac{9944 \pi ^4}{2025}
    -\frac{608201 \pi ^2}{2430}
  \biggr)
  +\delta ^6 
  \biggl(
    -\frac{284701}{540}
    +\frac{2624 a_4}{9}
    -\frac{1093 \pi ^2 \zeta _3}{15}
    \nonumber \\ &&
    +\frac{391 \zeta_3}{3}
    +\frac{772 \zeta _5}{3}
    +\frac{328 l_2^4}{27}
    -\frac{668}{135} \pi ^2 l_2^2
    -\frac{1484 \pi^2 l_2}{3}
    +\frac{4972 \pi ^4}{675}
    \nonumber \\ &&
    +\frac{591641 \pi ^2}{1620}
  \biggr)
  + \mathcal{O}(\delta^7 \ln^2(\delta))
  ~,
  \label{eq:X3}
\end{eqnarray}
where we specified the color factors to QCD, set $\mu_s=m_b$ and only show the first 
two expansion terms.
Furthermore we use the notations
$l_2 = \ln(2)$, $a_4 = \text{Li}_4(1/2)$ and $\zeta_i$ is 
Riemanns zeta function.
The full result expressed in terms of $SU(N)$ color factors and
up to $\mathcal{O}(\delta^{12})$ can be found in the 
ancillary file to Ref.~\cite{Fael:2020tow}.
Recently the results of three color factors up to $\mathcal{O}(\delta^9)$
have been confirmed in Ref.~\cite{Czakon:2021ybq}.

Analogously, we can give the result for $Q_1$:
\begin{eqnarray}
  Q_1 &=& \Gamma_0 m_b^2 \left( Y_0 
  + C_F \sum\limits_{i=1}^\infty \left( \frac{\alpha_s}{\pi} \right)^i Y_i
  \right)  
\end{eqnarray}
with $Y_0=3/10(1-\rho^{10})-9/2(1-\rho^6)\rho^2-24(1-\rho^2)\rho^4-18(1+\rho^2)\rho^4\ln(\rho^2)$.
Our result for the $\mathcal{O}(\alpha_s^3)$
correction reads:
\begin{eqnarray}
  Y_3 &=& 
  \delta ^7 
  \biggl(
    -\frac{52480 a_4}{567}
    +\frac{4372 \pi ^2\zeta _3}{189}
    -\frac{8188 \zeta _3}{189}
    -\frac{15440 \zeta _5}{189}
    -\frac{6560 l_2^4}{1701}
    +\frac{2672 \pi ^2 l_2^2}{1701}
    \nonumber \\ &&
    +\frac{89776 \pi ^2 l_2}{567}
    -\frac{19888 \pi^4}{8505}
    -\frac{608201 \pi ^2}{5103}
    +\frac{266929}{1701}
  \biggr)
  + \delta ^8 
  \biggl(
    \frac{26240 a_4}{189}
    -\frac{2186 \pi ^2 \zeta _3}{63}
    \nonumber \\ &&
    +\frac{3910 \zeta_3}{63}
    +\frac{7720 \zeta _5}{63}
    +\frac{3280 l_2^4}{567}
    -\frac{1336}{567} \pi ^2 l_2^2
    -\frac{2120 \pi ^2 l_2}{9}
    +\frac{9944 \pi ^4}{2835}
    +\frac{591641 \pi^2}{3402}
    \nonumber \\ &&
    -\frac{284701}{1134}
  \biggr)
  + \mathcal{O}(\delta^9 \ln^2(\delta))
  ~.
\end{eqnarray}
Since the leptonic momentum $q$ has to be ultrasoft,
i.e. $q \sim \delta \cdot m_b$, the $n$-th $q^2$ moment is 
suppressed by $2n$ additional powers of $\delta$
as compared to the leading $\delta^5$ term in Eq.~\ref{eq:X3}.

\begin{figure}[t]
  \centering
  \includegraphics[width=0.49\textwidth]{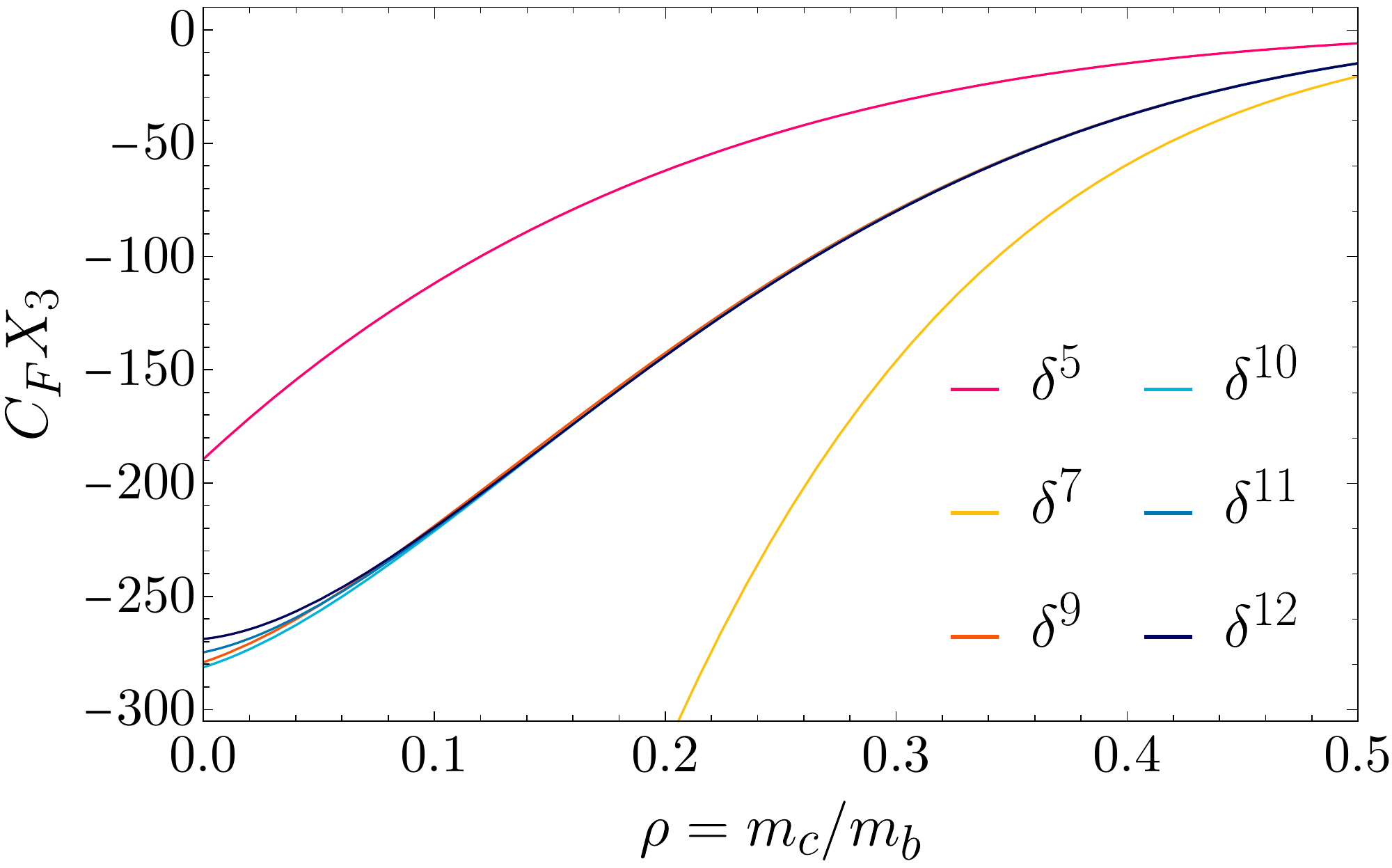}
  \includegraphics[width=0.49\textwidth]{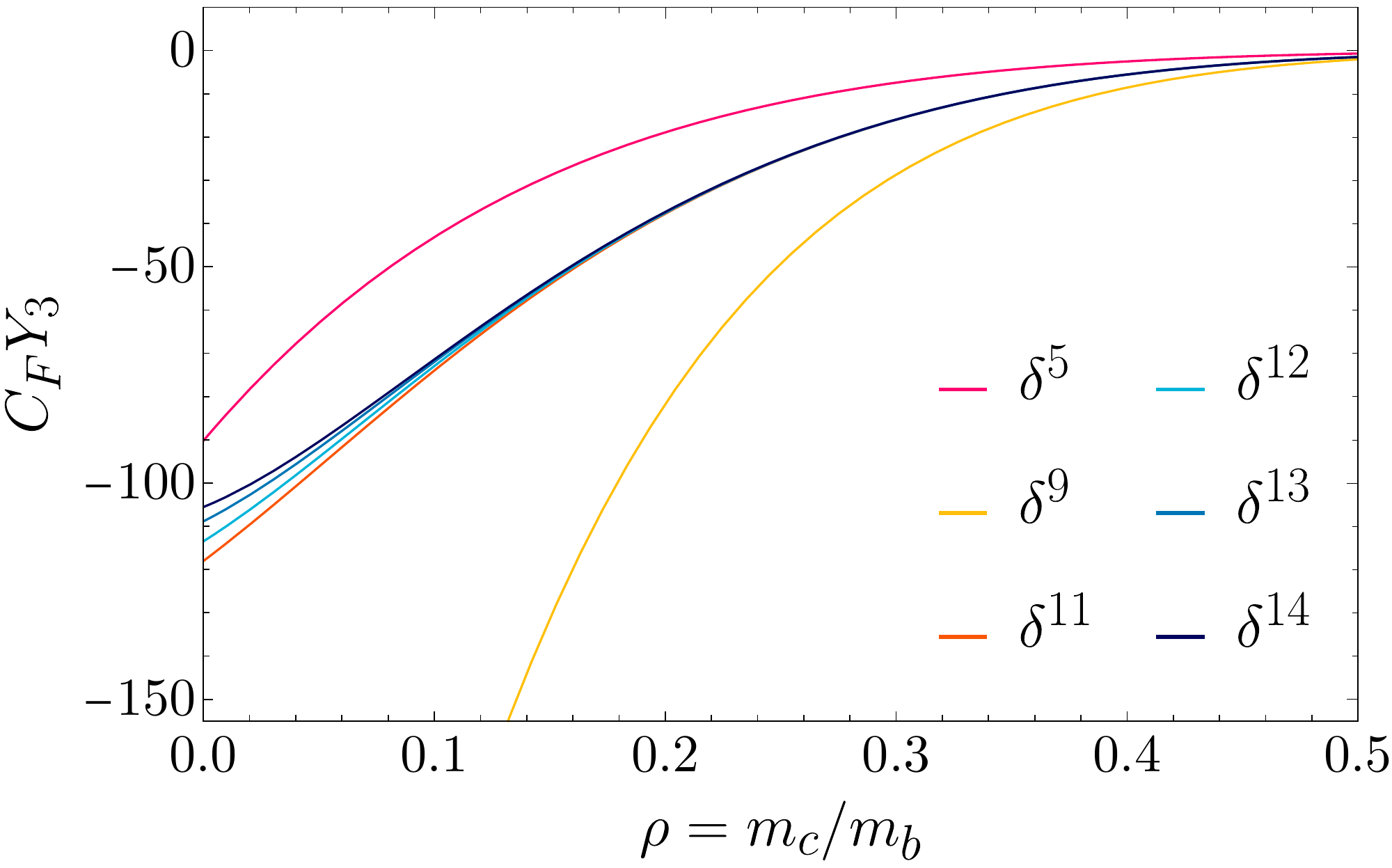}
  \caption{The $\mathcal{O}(\alpha_s^3)$ contribution to the total decay rate 
  $b \to s \ell \overline{\nu}$ (left) and its first $q^2$-moment (right)
  for different expansions depths in $\delta$.}
  \label{fig:1}
\end{figure}

The convergence of the $\delta$-expansion is studied in 
Fig.~\ref{fig:1}. 
For both, the total rate $\Gamma$ and the first $q^2$ moment
$Q_1$, one observes that the convergence at the physical point 
$\rho \sim 0.28$ is fast and does not vary much starting from order $\delta^{9}$
for the total decay rate and $\delta^{11}$ for $Q_1$.
Also at $\rho \to 0$ one sees a convergence for the high expansion 
terms, although much slower than at the physical point.
We get:
\begin{align}
  X_3(\rho=0.28) &= -68.4 \pm 0.3 
  ~, &
  Y_3(\rho=0.28) &= -14.41 \pm 0.03
  ~.
\end{align}
The uncertainty due to the truncation of the series has been determined 
from the difference of the last two expansion orders including a 
safety factor of $5$.
At 2-loop order this approach leads to a conservative error 
approximation.

\begin{figure}[t]
  \centering
  \includegraphics[width=0.49\textwidth]{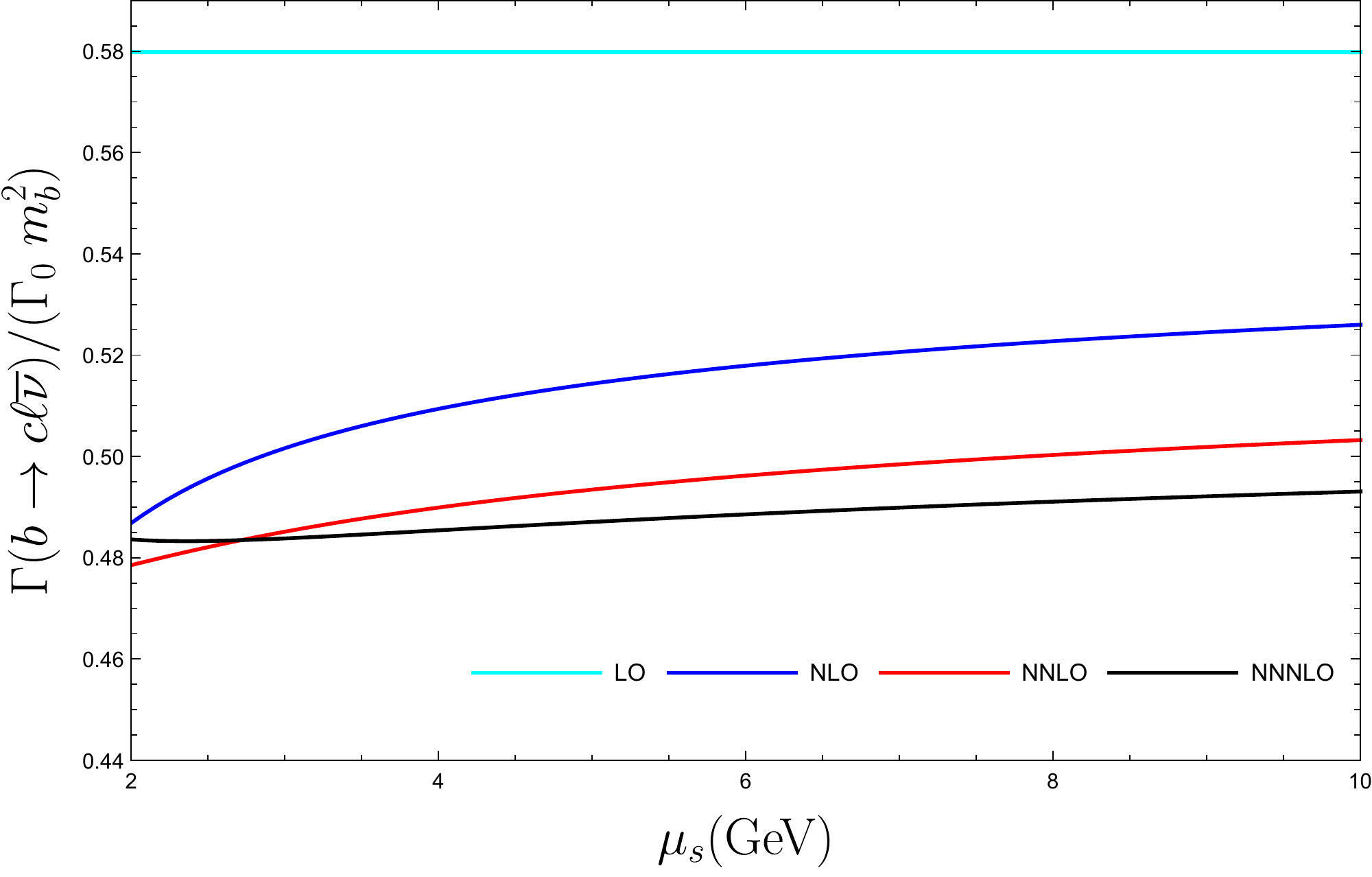}
  \includegraphics[width=0.49\textwidth]{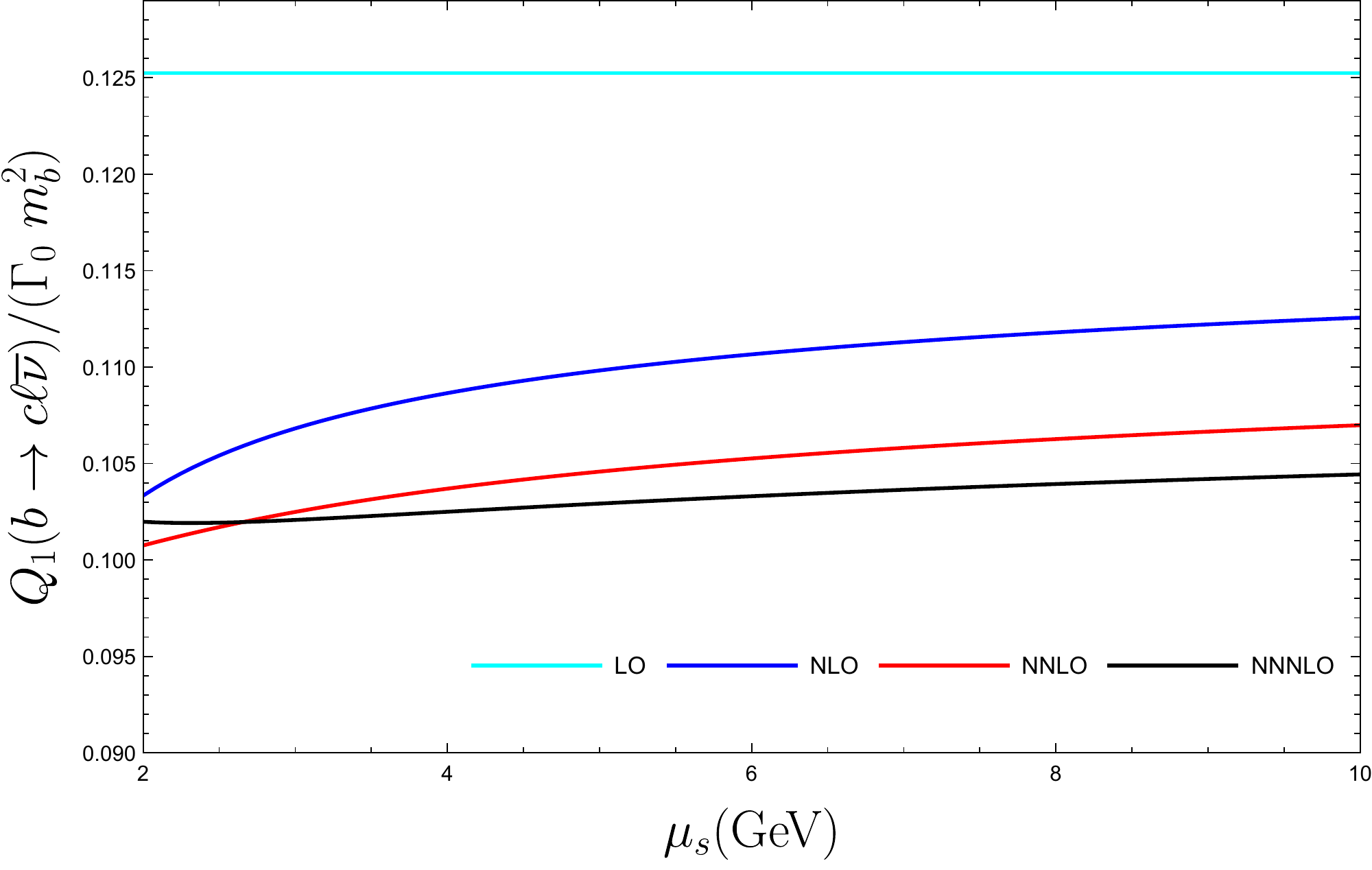}
  \caption{The dependence on $\mu_s$ of the total rate $\Gamma$ (left) and the
  first $q^2$-moment $Q_1$ (right) for different orders in the 
  strong coupling constants with the bottom quark expressed 
  in the kinetic scheme and the charm mass in the $\overline{\text{MS}}$ 
  scheme.
  The scale of the charm quark is set to $\mu_c=3\,\text{GeV}$.}
  \label{fig:2}
\end{figure}

Using the on-shell masses $m_c=1.3\,\text{GeV}$ and $m_b=4.7\,\text{GeV}$ and setting the 
renormalization scale $\mu_s = m_b$, we find
\begin{eqnarray}
  \Gamma(m_b,m_c) &=& \Gamma_0 X_0 \left[ 
      1 
    - 1.72 \frac{\alpha_s}{\pi} 
    - 13.09 \left( \frac{\alpha_s}{\pi} \right)^2 
    - 162.82 \left( \frac{\alpha_s}{\pi} \right)^3
    \right]
  ~,
  \\
  Q_1(m_b,m_c) &=& \Gamma_0 m_b^2 Y_0 \left[ 
    1 
  - 1.61 \frac{\alpha_s}{\pi} 
  - 12.83 \left( \frac{\alpha_s}{\pi} \right)^2 
  - 168.34 \left( \frac{\alpha_s}{\pi} \right)^3
  \right]
  ~.
\end{eqnarray}
As expected we find a bad convergence of the perturbative series using the 
on-shell scheme for the quark masses.
To mitigate this problem various so-called threshold masses have been 
proposed. 
We want to focus on the scheme used for the latest extraction of 
$|V_{cb}|$.
Here, the bottom mass is expressed in the kinetic scheme, while
the charm quark is expressed in the $\overline{\rm MS}$ scheme
at the scale $\mu_c=3\,\text{GeV}$.
With the input values 
$m_b^{\text{kin}}=4.526\,\text{GeV}$
and
$\overline{m}_c(3\,\text{GeV})=0.993\,\text{GeV}$,
we find 
\begin{eqnarray}
  \Gamma(m_b^{\text{kin}},\overline{m}_c(3\,\text{GeV})) &=& \Gamma_0 X_0 \left[
    1 
    - 1.67 \frac{\alpha_s^{(4)}}{\pi} 
    - 7.25 \left( \frac{\alpha_s^{(4)}}{\pi} \right)^2
    - 28.6 \left( \frac{\alpha_s^{(4)}}{\pi} \right)^3
  \right]
  ~,
  \\
  Q_1(m_b^{\text{kin}},\overline{m}_c(3\,\text{GeV})) &=& \Gamma_0 Y_0 \left(m_b^{\text{kin}}\right)^2 \left[ 
    1 
    - 1.83 \frac{\alpha_s^{(4)}}{\pi} 
    - 8.45 \left( \frac{\alpha_s^{(4)}}{\pi} \right)^2
    - 34.7 \left( \frac{\alpha_s^{(4)}}{\pi} \right)^3
    \right]
  ~,
  \nonumber \\
\end{eqnarray}
where $\mu_s=m_b^{\text{kin}}$ is used.
Note that the conversion to the kinetic scheme also contains the 
renormalization of the HQET parameters $\mu_\pi$ and $\rho_D$,
which formally only enter at order $1/m_b^2$ and $1/m_b^3$ respectively.
The scale dependence of the two quantities can be studied 
in Fig.~\ref{fig:2}. 
One observes a much better behavior of the perturbative series and 
a reduced dependence on the renormalization scale.

The results for the total cross section together with the improvement
of the relation between the on-shell and kinetic mass to $\mathcal{O}(\alpha_s^3)$
have recently been used to update the inclusive determination of 
$|V_{cb}|$ \cite{Bordone:2021oof}
\begin{eqnarray}
  |V_{cb}| = 42.16(30)_{th} (32)_{exp} (25)_\Gamma \times 10^{-3}
  ~.
\end{eqnarray}
The inclusion of the higher order calculations resulted in a small shift 
of the central value and a reduced theory uncertainty. 
Especially the uncertainty 
due to the width $\Gamma$ was halved.

The heavy daughter limit allows us also to estimate the $O(\alpha_s^3)$ correction to the $b \to u \ell \overline{\nu}_\ell$ decay
by setting $\delta \to 1$:
\begin{equation}
X_3^u = -202 \pm 20,
\end{equation}
where the relative 10\% uncertainty is estimated for unknown higher $\delta$ terms in the expansion.
We can study the convergence of the perturbative series also in this case.
We use the exact one and two loop results in the massless limit from Ref.~\cite{Pak:2008cp} 
and the three loop correction estimated above to derive the total rate for $b \to u \ell \overline{\nu}_\ell$
\begin{equation}
\Gamma_{b \to u} (m_b^\mathrm{kin}, \overline{m}_c(3 \text{ GeV})) =
\Gamma_0 X_0
\Bigg[
 1
 - 0.27 \frac{\alpha_s^{(4)}}{\pi}
 + 4.0 \left(\frac{\alpha_s^{(4)}}{\pi}\right)^2
 + 95.4 \left(\frac{\alpha_s^{(4)}}{\pi}\right)^3
\Bigg]~.
\end{equation}
We observe an apparent worse behavior of the $\alpha_s$ expansion compared to $b \to c$.
Note that the result depends in this case also on the Weak-Annihilation scale entering 
in the Wilson coefficient of $\rho_D$ at order
$1/m_b^3$. We set $\mu_\mathrm{WA} = m_b^\mathrm{kin}/2$.

If we specify the color factors to QED and set $\delta = 1 - m_e/m_\mu \approx 0.005$
we obtain a prediction for the muon lifetime $\tau_\mu$ via
\begin{eqnarray}
  \frac{1}{\tau_\mu} \equiv \Gamma(\mu^- \to e^- \nu_\mu \overline{\nu}_e)
  = \frac{G_F^2 m_\mu^5}{192 \pi^3} ( 1 + \Delta q) .
\end{eqnarray}
Precise measurements of the muon lifetime together with accurate  
QED predictions therefore allow the extraction of the Fermi 
constant $G_F$.
The various correction terms, see for example Ref.~\cite{vanRitbergen:1999fi} for 
a review, are usually parametrized via 
\begin{eqnarray}
  \Delta q = \sum\limits_{i \geq 0} \Delta q^{(i)} .
\end{eqnarray}
We find for the QED corrections
\begin{eqnarray}
  \Delta q^{(3)} \approx \left( \frac{\alpha(m_\mu)}{\pi} \right)^3 
  \left( -15.3 \pm 2.3 \right)
  ~,
\end{eqnarray}
where the error is estimated from the convergence properties 
at 1- and 2-loop order for which exact calculations are available
\cite{Kinoshita:1958ru,vanRitbergen:1998yd,Steinhauser:1999bx}.
This translates to a shift in the muon lifetime of \\
$\Delta \tau_\mu \approx (-9 \pm 1 )\times 10^{-8}\,\text{$\mu$s}$.
Comparing this with the current experimental value given by \\
$\tau_\mu = (2.1969811 \pm 0.0000022)\,\text{$\mu$s}$ \cite{Zyla:2020zbs}, 
we see that the new corrections are two orders of magnitude 
smaller than the current experimental uncertainty.
A new extraction of the Fermi constants therefore needs 
an improvement of the experimental data.

\section{Conclusions}
\label{sec:4}
In these proceedings we reviewed the calculation of the $\mathcal{O}(\alpha_s^3)$
corrections to the process $b \to c \ell \overline{\nu}$ retaining finite 
charm quark effects through an expansion around the equal mass case 
obtained in Ref.~\cite{Fael:2020tow}. 
Furthermore, we showed an extension of our method to inclusive $q^2$ moments,
which can be used to further constrain the global fits from which also 
$|V_{cb}|$ is extracted.
We showed that the expansions converge fast at the physical point and can 
even be extended down to $\delta \to 1$.
Although we find a badly converging prediction using the on-shell masses for 
charm and bottom, the predictions are improved by changing into the kinetic 
scheme for the bottom quark.
Since the knowledge of other moments, like moments of the 
lepton energy or the hadronic mass, is desireable for the 
global fits we plan to extend our calculation.
After specifying our results to QED we also obtain $\mathcal{O}(\alpha^3)$ 
predictions for the muon decay.

\section*{Acknowledgements}
I thank Matteo Fael and Matthias Steinhauser for their collaboration
on the presented work.
This research was supported by the Deutsche Forschungsgemeinschaft (DFG, German ResearchFoundation) 
under grant 396021762 — TRR 257 “Particle Physics Phenomenology after theHiggs Discovery”.

\bibliography{bibliography.bib}

\nolinenumbers

\end{document}